%% file: main-clean.tex
\documentclass[sigconf]{acmart}
\usepackage{threeparttable}
\usepackage{listings}
\usepackage{textcomp}
\usepackage[marginal]{footmisc}
\usepackage{latexsym}
\usepackage[utf8]{inputenc}
\usepackage{microtype}
\usepackage{array}
\usepackage{pifont}
\usepackage{tabularx}
\usepackage{adjustbox}
\usepackage{multirow}
\usepackage{tablefootnote}
\usepackage{enumerate}
\usepackage{xspace}
\usepackage{tcolorbox}
\usepackage{booktabs,dcolumn}

\usepackage{url}
\usepackage{amsmath,amsthm,bm,stmaryrd}

\usepackage{latexsym}
\usepackage{microtype}
\usepackage{soul}
\usepackage{booktabs}

\usepackage{algorithmicx}
\usepackage{algpseudocode}
\usepackage{algorithm}
\usepackage{caption}

\usepackage{float}
\usepackage{paralist}
\usepackage{xcolor}
\usepackage{color}
\usepackage{mathtools}
\usepackage{makecell}
\usepackage{multirow}
\usepackage{multicol}
\usepackage{array}
\usepackage{comment}
\usepackage{amsfonts}
\usepackage{textcomp}
\usepackage{fmtcount}
\usepackage{flushend}
\usepackage{thmtools}
\usepackage{xspace}
\usepackage{threeparttable}
\usepackage{colortbl}
\usepackage{subcaption}
\definecolor{mygray}{gray}{.9}
\usepackage{fixltx2e}

\usepackage[inkscapelatex=false]{svg}
\usepackage{amstext} 
\usepackage{array}   
\newcolumntype{L}{>{$}l<{$}}

\AtBeginDocument{%
  }

\copyrightyear{2024}
\acmYear{2024}
\setcopyright{acmlicensed}
\acmConference[CIKM '24]{Proceedings of the 33rd ACM International Conference on Information and Knowledge Management}{October 21--25, 2024}{Boise, ID, USA}
\acmBooktitle{Proceedings of the 33rd ACM International Conference on Information and Knowledge Management (CIKM '24), October 21--25, 2024, Boise, ID, USA}
\acmDOI{10.1145/3627673.3680016}
\acmISBN{979-8-4007-0436-9/24/10}
\begin{document}


\title{RCAgent: Cloud Root Cause Analysis by Autonomous Agents with Tool-Augmented Large Language Models}

\author{Zefan~Wang*}
\affiliation{%
  \institution{Tsinghua University}
  \city{Beijing}
  \country{China}
}
\email{wang-zf20@mails.tsinghua.edu.cn}

\author{Zichuan~Liu*}
\affiliation{%
  \institution{Nanjing University}
  \city{Nanjing}
  \country{China}
}
\email{zichuanliu@smail.nju.edu.cn}

\author{Yingying~Zhang†}
\affiliation{%
  \institution{Alibaba Group}
  \city{Hangzhou}
  \country{China}
}
\email{congrong.zyy@alibaba-inc.com}

\author{Aoxiao~Zhong}
\affiliation{%
  \institution{Havard Univerrsity}
  \city{Cambridge}
  \country{USA}
}
\email{aoxiaozhong@g.harvard.edu}

\author{Jihong~Wang*}
\affiliation{%
  \institution{Xi'an Jiaotong University}
  \city{Xi'an}
  \country{China}
}
\email{wang1946456505@stu.xjtu.edu.cn}

\author{Fengbin~Yin}
\affiliation{%
  \institution{Alibaba Group}
  \city{Hangzhou}
  \country{China}
}
\email{yinfengbin.yfb@alibaba-inc.com}

\author{Lunting~Fan}
\affiliation{%
  \institution{Alibaba Group}
  \city{Hangzhou}
  \country{China}
}
\email{lunting.fan@taobao.com}

\author{Lingfei~Wu}
\affiliation{%
  \institution{Anytime AI}
  \city{New York}
  \country{USA}
}
\email{lwu@anytime-ai.com}

\author{Qingsong~Wen†}
\affiliation{%
  \institution{Squirrel Ai Learning}
  \city{Bellevue}
  \country{USA}
}
\email{qingsongedu@gmail.com}

\thanks{$\star$
This research was primarily done during the internship at Alibaba Group.
 
$\dagger$ Corresponding authors. 
}

\renewcommand{\shortauthors}{Zefan Wang et al.}

\begin{abstract}

Large language model~(LLM) applications in cloud root cause analysis~(RCA) have been actively explored recently. 
However, current methods are still reliant on manual workflow settings and do not unleash LLMs' decision-making and environment interaction capabilities. 
We present RCAgent, a tool-augmented LLM autonomous agent framework for practical and privacy-aware industrial RCA usage. 
Running on an internally deployed model rather than GPT families, RCAgent is capable of free-form data collection and comprehensive analysis with tools. 
Our framework combines a variety of enhancements, including a unique Self-Consistency for action trajectories, and a suite of methods for context management, stabilization, and importing domain knowledge. 
Our experiments show RCAgent's evident and consistent superiority over ReAct across all aspects of RCA—predicting root causes, solutions, evidence, and responsibilities—and tasks covered or uncovered by current rules, as validated by both automated metrics and human evaluations. 
Furthermore, RCAgent has already been integrated into the diagnosis and issue discovery workflow of the Real-time Compute Platform for Apache Flink of Alibaba Cloud. 
\end{abstract}



\maketitle

\vspace{-2mm}
\section{Introduction}
Cloud computing platforms have been increasingly utilized for application and service deployment in recent years~\cite{chen2019empirical, ma2022empirical}. 
Anomalies in cloud computing systems, such as unrecoverable failures and hanged jobs, severely impact customer experience and can potentially violate service level agreements\cite{zhang2022tfad,ahmed2023recommending}. 
Root Cause Analysis~(RCA)~\cite{zhang2021cloudrca, nguyen2013fchain, aggarwal2020localization}, a core component of site reliability engineering, is currently receiving ongoing attention from large cloud computing enterprises such as Amazon, Microsoft, Google, and Alibaba.
To increase the efficiency of cloud service reliability enhancement, a series of Artificial Intelligence for Operations~(AIOps) approaches~\cite{chen2016causeinfer,wang2020root,zhang2022netrca}
have been widely adopted in RCA to reduce the MTTR~(mean time to resolve).
While these typical AIOps aid in automated processes, their application faces challenges such as poor data quality, shifting data distribution, laborious data annotation, and limited generalization for models~\cite{cheng2023ai}. 

The advancements in Large Language Models~(LLMs), especially within the GPT~\cite{brown2020language,ouyang2022training,openaichatgptblog,openai2023gpt4} and LLaMA~\cite{touvron2023llama,touvron2023llama2} families, indicate an intriguing future of solving intricate reasoning tasks.
Recent works demonstrate the use of LLMs in cloud RCA tasks. 
~\cite{ahmed2023recommending,jin2023assess} fine-tune GPT models for root causes summarization. 
These works rely heavily on the computationally expensive supervised adaption to cloud system tasks and do not fully utilize the generalization and reasoning abilities of LLMs.
One possible solution is to use few-shot RAG~\cite{lewis2020retrieval} on LLMs, with representative methods such as RCACopilot~\cite{chen2023empowering}, PACE-LM~\cite{zhang2023pace}, and Xpert\cite{jiang2023xpert}. 
However, these works are all based on the GPT family and scenarios within Microsoft, not addressing the data privacy concerns associated with using LLMs with cloud system data.
Furthermore, none of the above methods leverage the autonomous capabilities of LLMs for information collection, decision-making, and environmental interaction~\cite{yao2022react}.

Tool-augmented autonomous agents, as demonstrated in early experiments~\cite{bubeck2023sparks}, further unlock the potential of LLMs in interactive environments. 
By equipping LLMs with defined tools and associated documentation, and by facilitating tool invocation through mechanisms like function calls or command line inputs, and then executing these tools and returning environmental feedback, LLMs can handle tasks that require extensive expertise and abilities. 
A representative paradigm within the realm of autonomous agents is ReAct~\cite{yao2022react}, a workflow that embodies a thought-action-observation loop and offers flexibility for extensions~\cite{liu2023bolaa}. 
However, the adoption of LLM agents in the AIOps field, especially with noisy and lengthy data, remains limited~\cite{le2022log, locke2021logassist}.
The primary challenges are action validity and context length, both of which heighten the demands of LLM-as-agent capabilities~\cite{liu2023agentbench}. 
Also, to the best of our knowledge, there is no interactive environment built upon realistic production-level RCA problems for LLM agents to operate on. 

To this end, we introduce \textbf{RCAgent}, the first practical LLM-based RCA framework within the tool-augmented autonomous agent paradigm. 
We design an enhanced prompting cycle skeleton and an interactive environment enriched with external knowledge and stabilization techniques, tailored for LLM agents to handle diverse data types.
Additionally, we design aggregation methods for action trajectories and text output, combining suboptimal results from LLMs. 
Unlike ReAct, our approach operates in a trajectory-level zero-shot way, eliminating the need for manual or auto-generated action examples. 
Furthermore, to facilitate general and secure industrial usage, we forgo the use of powerful external API models
and implement this framework on a locally deployed model, underscoring the efficacy of our stabilization method.

The analysis results from RCAgent are being utilized in the Real-time Compute Platform for Apache Flink of Alibaba Cloud to diagnose anomalous stream processing jobs uncovered by current methods. 
We have incorporated a feedback mechanism in the company to identify issues in the PaaS and IaaS layers of the cloud system, offering insights for development teams. 

We summarize our contributions as follows: 
\begin{itemize}
\item We propose RCAgent, the first tool-augmented agent based on LLM for privacy-aware real-world cloud RCA, unleashing the decision-making ability of LLMs in the AIOps field. 
\item We introduce a bag of methods to enhance the tool agent, including aspects of prompting framework, tool setting, stabilization, and aggregation methods. These make the agent based on locally deployed LLM a valid solution for complex environments like cloud systems. 
\item We demonstrate the practical usage of RCAgent with real-world experiments on computing jobs in Alibaba Cloud. 
\end{itemize}

\section{Challenge}
Though tool-augmented LLM agents provide new possibilities for the cloud RCA task, including autonomous decision-making and handling unseen anomalies without laboriously annotated training data, several critical challenges exist. 

\paragraph{Privacy}
A general LLM method for RCA and other cloud AIOps tasks should be internally hosted for security concerns. 
Specifically, transmitting production-level confidential data to external API induces privacy risks. 
This means stronger models like ChatGPT cannot be used except for those close collaboration enterprises. 
While trading off the model's ability for security, we need techniques to mitigate the gap to some extent.

\paragraph{Context Length}
\label{challenge:context length}
A fundamental problem for agent usage in realistic cloud environments is context length because various kinds of data, such as logs, code, and database query results, tend to be enormous. 
Even if the LLMs can extrapolate to larger context length\cite{pal2023giraffe}, processing unnecessarily excessive tokens is highly inefficient.  

\paragraph{Action Validity}
Open-ended action generation for LLMs is of great challenge because less sufficiently aligned LLMs have a larger possibility of generating invalid actions\cite{liu2023agentbench}. 
These errors severely damage the performance of autonomous agents on comprehensive tasks.
The model restriction from privacy concerns and noisy cloud data this problem even more arduous.

\section{Methodology}

\begin{figure}[!tb]
    \centering
    \includegraphics[width=.95\linewidth, trim=350 0 0 0,clip]{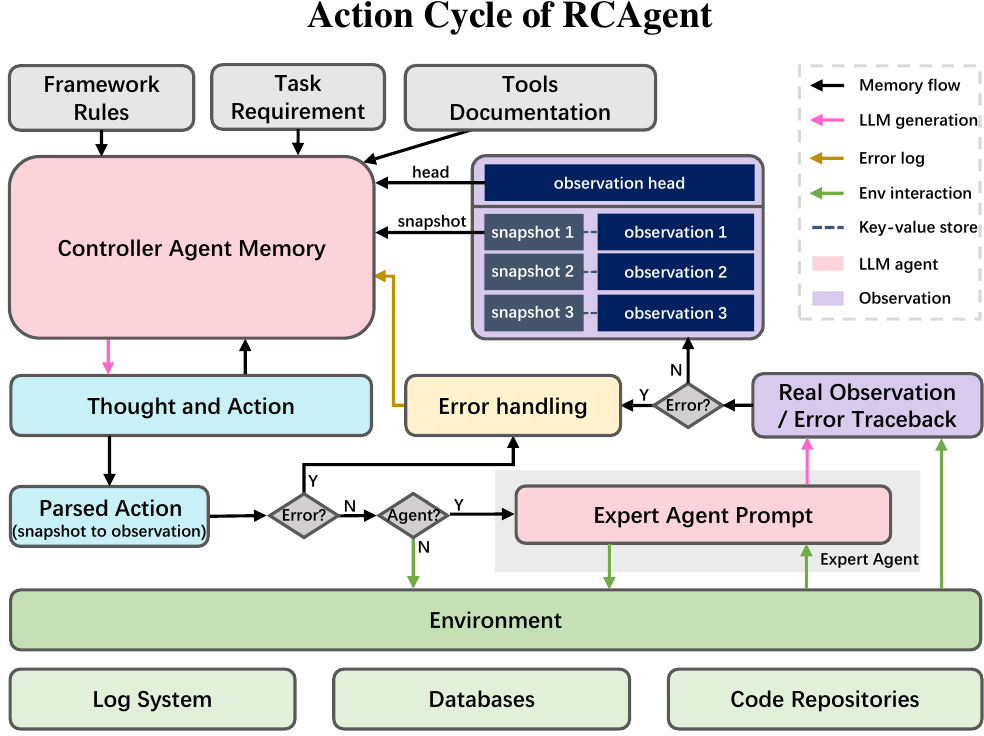}\vspace{-2mm}
    \caption{
        Overview of the action cycles from RCAgent.
        The cycle involves generating verbal thoughts, taking actions, and receiving observation from the environment, all of which are recorded in the prompt alongside the initial memory to boost reasoning.
        Besides, RCAgent includes the key-value store for observation retrieval, allowing the agent to operate on lengthy text data.
        After parsing the action, RCAgent executes it directly or invokes an expert agent, depending on its type. 
    }\vspace{-3mm}
    \label{fig:overview}
\end{figure}

To systematically and reliably prompt the LLM as a tool-augmented autonomous agent for cloud RCA, we propose \textbf{RCAgent}, an enhanced reasoning and acting framework. 
An overview of our methodology at the decision-loop level is shown in Figure~\ref{fig:overview}. 
For disambiguousity, the LLM agent with the prompt of thought-action-observation loop is named the \textit{controller agent} responsible for coordinating actions, and RCAgent additionally employs the LLM as tools called the \textit{expert agents} for domain-specific functionalities. 

In concordance with the typical implementation of ReAct 
tool agent prompt framework~\cite{qin2023toolllm}, the controller agent is injected with three basic prompts: 
(i) framework rules that describe the thought-action-observation loop, 
(ii) task requirements that contain instructions for the RCA tasks with basic cloud knowledge, and 
(iii) tools documentation that describes the description of all invokable tools.
Because of its flexibility and readability, JSON is chosen as the data interchange format for all generations in the action step from LLM. 
We also define a tool named `finalize' as an exit point that allows the model to freely decide when to report findings in a parsable format.
Note that RCAgent discards the few-shot examples compared to the original ReAct because of limited context length. 

Starting from the tool agent version of ReAct, we propose several enhancements to address the challenges of using tool-augmented LLM agents in cloud RCA. 
To deal with the context length challenge described in \S~\ref{challenge:context length}, we first invent an observation management method for compressing token usage introduced in \S~\ref{method:obsk}. 
Then, because the LLM itself does not have access to and enough domain-specific knowledge about the cloud system, we build tools including LLM-augmented ones for the RCAgent, whose designs are described in \S~\ref{method:tools}. 
Addressing the action validity problem, RCAgent has stabilizing methods presented in \S~\ref{method:stablize}. 
Lastly, to improve the performance of RCAgent since we are using less capable locally hosted LLM, we utilize the aggregation method in RCAgent described in \S~\ref{method:sc}. 

\vspace{-1mm}
\subsection{Observation Snapshot Key}\label{method:obsk}
One of the basic challenges of building autonomous agents in a comprehensive cloud environment is context length~\cite{chen2023empowering}. 
The most inflating part of the agent prompt is the observation content in the action trajectories, containing a large amount of logs, table entries, etc. 
To overcome the information loss from truncating and summarizing observations, we propose OBservation Snapshot Key~(\textbf{OBSK}), a new method to address the context length problem in realistic cloud tasks. 
As shown in Figure~\ref{fig:overview}, OBSK only shows the head of observation to the controller agent, leaving a hash ID (snapshot key) for further usage. A key-value store is built for mapping the snapshot key to real observation. 
Thus, when a snapshot key is found in a parsed action, RCAgent queries through the key-value store and returns the corresponding observation. 
This ensures necessary information with controlled length is provided for the controller agent as supportive information for decision-making. 

\vspace{-1mm}
\subsection{Tool Preparation}
\label{method:tools}
We employ data querying functions as \textit{information-gathering tools} and LLM-based expert agents as \textit{analytical tools}, similar to the data collection and analysis process done by human SREs. 

\vspace{-1mm}
\subsubsection{Information-gathering Tools} 
Information-gathering tools are designed in an easy-to-use way, hiding all unrelated details in accessing data in cloud systems. 
For example, instead of giving SQL interface and Log query API to LLM, these tools only accept simple parameters like the ID of entities. 
This semantically minimalist tool setting will reduce the threshold for LLMs to take valid actions,  preventing useless exploration in large data warehouses. 
Also, to avoid the agent's ineffective analysis, we deduplicate similar information and exclude messages beneath the WARNING level.

\begin{figure}[tb!]
\vspace{-2mm}
    \centering
    \includegraphics[width=0.75\linewidth]{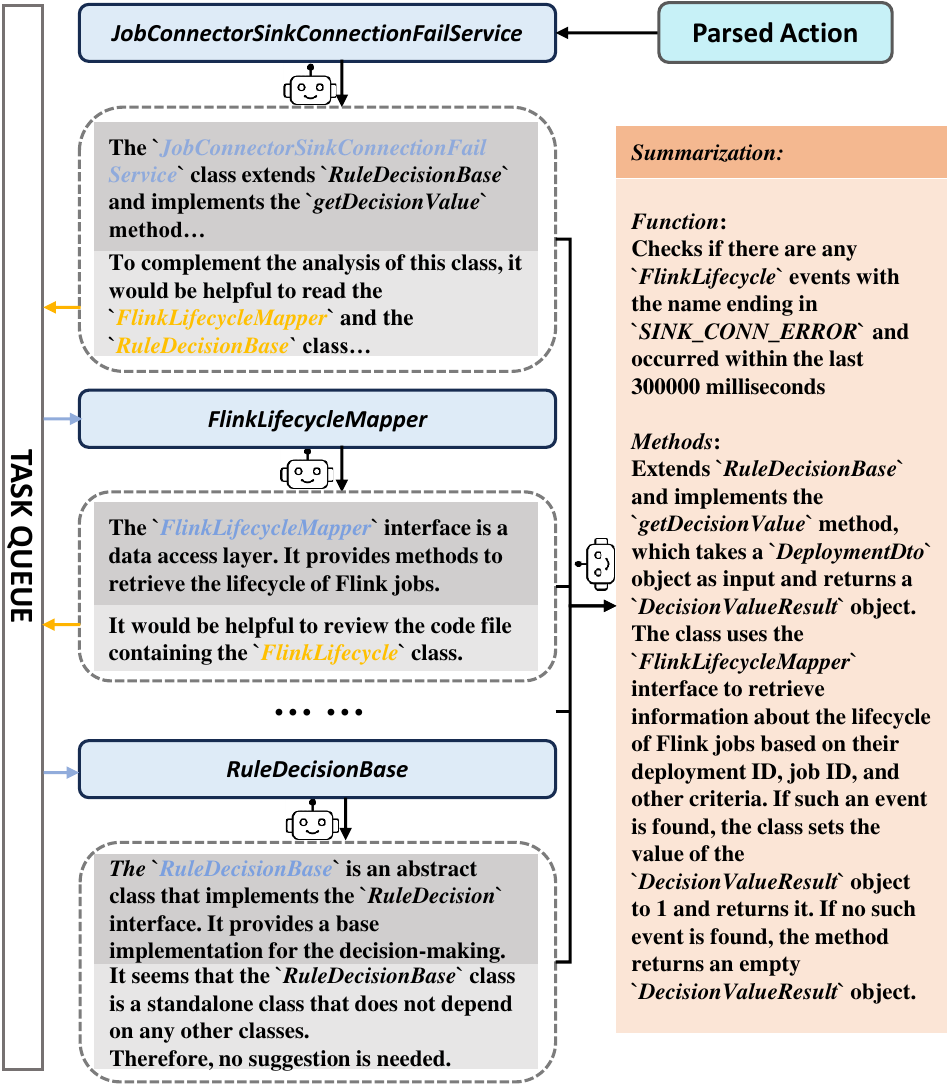}\vspace{-2mm}
    \caption{
    Code analysis tool in RCAgent.
    }\vspace{-6mm}
    \label{fig:codeagent}
\end{figure}

\vspace{-1mm}
\subsubsection{Analytical Tools} 

Analytical tools are proposed to extend the domain knowledge and abilities of the controller agent
augmented by LLMs with their reasoning ability. 
We name this kind of analytical tool the expert agent, which is shown in Figure~\ref{fig:overview}. 
We provide two expert agents for RCAgent as complementary knowledge tools, called the \textit{Code analysis tool} and the \textit{Log analysis tool}. 
Both generate analyses and aggregations prompted by the zero-shot Chain-of-Thought~(CoT)~\cite{kojima2022large} and answer extraction instructions. 

\vspace{-1mm}
\paragraph{Code analysis tool.} 

The code analysis tool works in a recursive manner, which is shown in Figure~\ref{fig:codeagent}. 
Given a class name, the code analysis tool searches the corresponding file in the code repository. 
After the LLM reading and analyzing the code file, it is prompted to suggest any other classes that would be helpful to analyze as supportive information. 
These suggestions from each code-reading round will be stored in a task queue, managing all pending tasks. 
With this exhaustive search, the code analysis tool stops parsing when no more code files of interest are recommended, or when all remaining recommended files are external dependencies.
Then, we utilize an LLM to summarize all the code files, whose result is presented to the controller agent as the observation. 

\begin{algorithm}[!tb]

 \caption{Pseudo code for log expert agent. }
 \label{alg:logagent}
 \begin{algorithmic}[1]

\Require Log $L$, Max prompt length $N$
\Ensure Interpretations $\tilde{R}$, Evidences $\tilde{E}$ 

    \State $S \gets$ split $L$ using delimiters (e.g., newline)
    \State $\mathbf{v}_s \gets$ \Call{EmbeddingModel}{$s$} for each $s$ in $S$
    \State $W=\{w_{ij}\}$ empty weight matrix
    \For{pairs $(s_i, s_j)$ in $S \times S$ where $j-i \in (0, 200]$}
        \State $d_{ij} \gets$ position distance between $s_i$ and $s_j$ in $L$
        \State $w_{ij} \gets CosSim(\mathbf{v}_{s_i}, \mathbf{v}_{s_j}) \times exp(-d_{ij})$
    \EndFor
    \State $C \gets$ \Call{LouvainClustering}{$S, W$}
    \State $C' \gets$ \Call{GreedyOverlapRemoval}{$C$}
    \State $P \gets$ partitions from $L$ indicted by components $C'$
    \State $R', E' \gets$ empty initialized filtered results
    \For{each partition $p$ in $P$}
        \State $E, A \gets$ retrieved sorted examples and answers
        \State $ICP \gets E[0:N], A[0:N]$ \Comment{In-context prompt}
        \State $R, E \gets$ \Call{LLMAnalysis}{$ICP, p$} 
        
        \For{each $(r, e)$ in $R,E$}
            \If{\Call{Levenshtein}{$e, p$} $<$ \Call{L}{$p$} - \Call{L}{$e$} $\times$ 0.9}  
                \State $R', E'\gets R' \cup r,  E' \cup e$
                \Comment{Filter hallucinations}
            \EndIf
        \EndFor
        
    \EndFor
    
    \State $\tilde{R}, \tilde{E} \gets$ \Call{LLMSummary}{$R', E'$}
 \end{algorithmic} 
 \end{algorithm}

\vspace{-1mm}
\paragraph{Log analysis tool.} 
The log analysis tool operates in an in-context RAG paradigm with some adaptions to lengthy log data. 
The complete mechanism is shown in Algorithm~\ref{alg:logagent}. 
We split the log $L$ into lines $S$ and built edges between lines with the cosine similarity of embedding exponentially decayed by document distance as weights $W$. 
This yields a weighted undirected dense graph $(S, W)$ regarding lines as vertices, describing the relevance of lines that are weakened when other lines are inserted in the middle. 
Then the graph is clustered with Louvain community detection~\cite{blondel2008louvain}, and the overlaps between clusters are removed by greedily switching the minimum amount of clustering labels. 
This clustering functions as semantic partitioning, and the result log chunks $P$ are then fed into the log agent one chunk per round to perform Retrieval-Augmented Generation~(RAG)~\cite{lewis2020retrieval}. 
Moreover, we instruct the expert agent to output evidence supporting its analysis by directly copying log content, preventing hallucinations of analyzing examples rather than the partitioned chunk. 
If the evidence listed by LLMs cannot be fuzzy-matched to the chunk $p$, the analysis result is discarded. 
Thus, we ensure reliable RAG on lengthy non-natural language. 

\vspace{-2mm}
\subsection{Stabilization}\label{method:stablize}
To overcome the degradation of action validity induced by noisy data and less capable local LLMs, we introduce two stabilizations. 

\vspace{-2mm}
\subsubsection{JSON Repairing}
\label{sec:json}
One of the vital problems in real-world applications of tool-augmented LLM autonomous agents is structured inference for parsable data. 
To our knowledge, there is no pain-free method to guarantee a specific data format (e.g., JSON) for interactions between LLM agents and the environment. 
Even though there are some structure helper toolkits, such as Outlines~\cite{willard2023cfg2} and TypeChat~\cite{typechat}, 
they either cannot generate free-form JSON with extensive escape characters while not impair generation quality, or solely rely on LLMs' capability of token error correction. 
To solve this issue, we employ an intuitive and effective method to generate structured interchange data named \textbf{JsonRegen}. 

Before LLM inference, all sensitive characters that may correspond to control symbols in JSON are replaced with insensitive ones for a clean prompt. 
When trivial cleaning 
fails to make the JSON-like string
parsable, a regeneration process is performed. 
To enforce the LLM to understand JSON structure, we instruct it to convert the content to YAML. 
The LLM is then prompted to regenerate a JSON with the same structure and content. 
The regeneration proceeds for several rounds or until a valid JSON is parsed. 

\subsubsection{Error Handling}
\label{sec:error}
The previous work~\cite{qin2023toolllm} demonstrates that LLMs in tool invocation tend to propagate errors, limiting exploratory actions. 
These issues are even more pronounced in less capable LLMs. 
Inspired by \cite{shinn2023reflexion}, we use pre-defined criteria to mark problematic actions or states as erroneous. 
As shown in Figure~\ref{fig:overview}, we provide error messages and suggestions to the controller agent, including these circumstances: 
(i) duplicate invocation of stateless tools with the same set of arguments, (ii) trivial input to expert agents, 
and (iii) early finalizing without thorough investigation. 
These error messages can reduce the frequency of meaningless actions taken by the control agent by alerting it. 

\vspace{-2mm}
\subsection{Self-Consistency Aggregation}
\label{method:sc}
Self-Consistency~(SC)~\cite{wang2022self} has proved its efficacy in various close-ended NLP tasks while aggregating sampled open-ended multi-step generation like RCA with LLM agent is underexplored. 
To our knowledge, utilizing SC on ReAct style trajectories is also not well-defined. 
Thus, we propose applying the SC paradigm to free-form generation on the topic of LLM autonomous agents.
\vspace{-1mm}
\subsubsection{Self-Consistency for Text Data.} 
To apply SC to text data, we utilize two methods in our experiments: 
\vspace{-1mm}
\paragraph{Vote with embedding}. We directly generalize the idea of unweighted SC~(majority vote), which performs best across all tasks~\cite{wang2022self}. 
The voting can be rewritten as
$\mathbf{arg\,max}_i (\operatorname{Sim}(\mathbf{a}_i, \frac{1}{K}\sum_j^K(\mathbf{a}_j))),$ where $K$ is sample count, and $\mathbf{a}_i$ is a one-hot vector representing sampled result $i$ with each position as a candidate choice or numerical result. 
We simply replace $\mathbf{a}$ as semantic embeddings for text output. 
This intuitively means the text result closest to the majority is chosen.
\vspace{-1mm}
\paragraph{Aggregate with LLMs}. 
Considering the possible diversity of generated content, we prompt LLM to aggregate the candidates and output in similar form and length. 

\begin{figure}[!tb]
    \centering
    \includegraphics[width=0.850\linewidth,trim=320 0 0 0,clip]{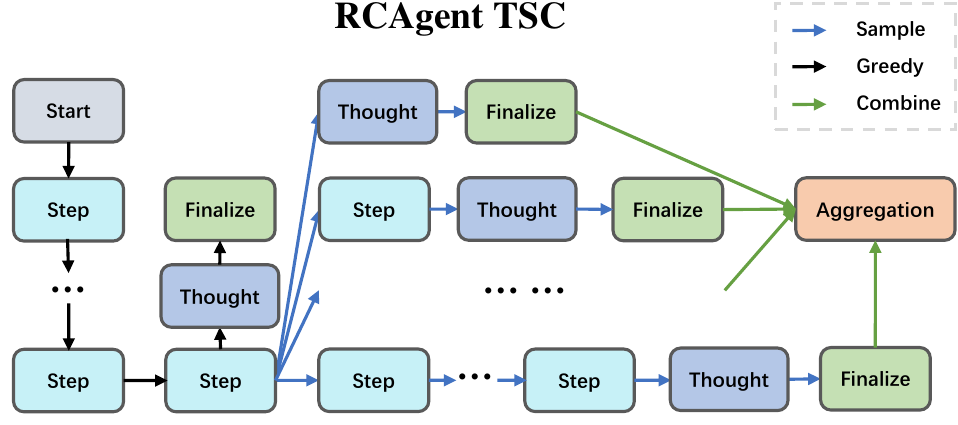}\vspace{-1mm}
    \caption{
    Trajectory-level Self-Consistency. Every \textbf{Step} in RCAgent means a sequential procedure of thought, action, and observation. 
    }\vspace{-5mm}
    \label{fig:selfconsistency}
\end{figure}

\vspace{-1mm}
\subsubsection{Self-Consistency for Tool Using Trajectories} 
SC has been comprehensively tested on CoT reasoning paths, and can naturally be utilized on ReAct style trajectories. 
However, directly sampling multiple cycles of thought-action-observation can be expensive. 
This is even more costly while some actions, such as activating expert agents, require heavy consumption. 
Moreover, random sampling without any history actions or few-shot examples leads to flooding errors, e.g. consecutive calling non-existent tools. 

Therefore, we propose a mid-way sampling method named Trajectory\nobreakdash-level Self-Consistency~(\textbf{TSC}) as shown in Figure~\ref{fig:selfconsistency}. 
Specifically, only when the controller agent is stepping into finalization does the sampling start.
This sampling strategy shares most preliminary steps between trajectory samples and reduces unnecessary consumption. 
Besides, the more stable action history from greedy decoding provides exemplification without additional context-length consumption from few-shot examples, suppressing the validity drop from sampling. 
This method strikes a balance between full-process SC on agent trajectories and one-step CoT SC. 

\vspace{-1mm}
\section{Experiment}

We develop and evaluate RCAgent on the Real-time Compute Platform in Alibaba Cloud, an enterprise-level and high-performance system capable of real-time stream data computation based on Apache Flink. 
This system achieves a throughput of $100$ million data records per second during peak hours.

\vspace{-1mm}
\subsection{Model Configuration}
Our implementation is based on \textbf{Vicuna-13B-V1.5-16K}~\cite{zheng2023judging} with vLLM~\cite{kwon2023efficient} backend on a single NVIDIA A100 SXM4 GPU (80 GB). 
We use the greedy decoding strategy by default for better reproducibility and stability. 
During self-consistency where random sampling is required, we use the default configuration of Vicuna.

The embedding model we use is \textbf{GTE-LARGE}~\cite{li2023towards}, for its slightly better results on MTEB~\cite{muennighoff2022mteb} than text-embedding-ada-002, providing an internally deployable substitute. 

For Self-Consistency results, we use $10$ output samples by default. 
We employ a step-wise Self-Consistency denoted as \textbf{SC}, which only accepts samples that finalize synchronously after the greedy decoding trajectory, allowing no additional action steps. 
\vspace{-1mm}
\subsection{Dataset Preparation}
\subsubsection{Anomaly Selection}
For root cause analysis, we collect a dataset of $15,616$ anomalous jobs of one-month cloud system history, either unrecoverable fail, or fail to start in $6$ minutes. 
We filter the data and obtain \textasciitilde$5,000$ non-trivial anomalous jobs with substantial log content. 
We use the Flink Advisor knowledge base, which is a large rule set distilled from experienced SREs' domain knowledge, to create analysis results for these jobs. 
Due to the imbalance of anomalies, which means a large proportion of anomalies have the same root cause, 
we reduce the successfully analyzed jobs to an offline dataset of $161$ jobs.
The reduction is done with the class-balance constraint that no more than two jobs have identical root causes.
The required annotation of these jobs contains four items:
the root cause, solution, evidence, and responsibility determination. 
We first use LLM to summarize the analysis from Flink Advisor and output the above four items. 
Then the SRE team proofreads and annotates the target output.

We guarantee that our annotations do not show uninformative patterns like ``The root cause of this anomaly is \dots{}'' that cause some of the semantic scores untrustworthy discovered in~\cite{jin2023assess,ahmed2023recommending}.

\vspace{-1mm}
\subsubsection{Data Sources}
The available data sources, on which we build information-gathering tools for the LLM agent, include:
\begin{itemize}
\item Log data at three levels: platform, runtime, and infrastructure, stored in SLS~(Simple Log Service) of Alibaba Cloud. 
\item Database containing the history of advisor services
\item Repositories containing the code of advisor services
\end{itemize}

For log and database entries, only data before the detection time of the anomaly can be retrieved, preventing the analysis of future information and adhering to real-world usage. 

The retrieval log database for the expert agent is a history subset of Flink Advisor with no overlap with the content used for labeling.

\subsection{Evaluation Metrics}
Besides semantic metric scores including \textbf{METEOR}~\cite{banerjee2005meteor}, 
\textbf{NUBIA}~\cite{kane2020nubia} (6-dim), \textbf{BLEURT}~\cite{sellam2020bleurt}, and \textbf{BARTScore}~\cite{yuan2021bartscore} (F-Score, CNNDM), we use additional embedding Score~(\textbf{EmbScore}), the cosine similarity from the default embedding model in our experiment.  

We also follow the common practice of using stronger models to estimate model prediction~\cite{zheng2023judging,qin2023toolllm,gao2023llama}. 
We use greedy decoding \textbf{gpt-4-0613}, a frozen version of GPT4\cite{openai2023gpt4} for better reproducibility. 
We prompt the model to judge the accuracy and helpfulness of root cause and solution predictions, marked as \textbf{G-Correctness} and \textbf{G-Helpfulness}, respectively, and give a score within $0\sim 10$.

\section{Result}

\subsection{Effectiveness}
We present the effectiveness of RCAgent on the offline dataset in Table~\ref{tab:root_cause_main}, \ref{tab:solution_main}, and~\ref{tab:evidence_main}. 
RCAgent outperforms the original ReAct in all aspects of comprehensive RCA encompassing root cause, solution, and evidence prediction. 
The performance superiority is evident and consistent across all metrics, including $+8.71$ and $+6.52$ METEOR against ReAct in the root cause and solution prediction subtasks.

Employing TSC aggregation using LLM summarization, the overall performance of RCAgent gains further enhancements, especially on solution prediction, witnessing gains of $+3.51$ METEOR, $+4.50$ BLEURT, and $+2.28\%$ G-Helpfulness. 
This boost can be explained by the broader diversity of solution sampling.

\input{table/rootcause}
\input{table/solution}
\input{table/evidence}

\subsection{Ablation Study}
To gauge the contribution of each component of RCAgent, we conduct an ablation study by removing enhancements introduced by RCAgent, including LLM expert agents, JsonRegen, and OBSK. 
The ablative result is shown in Table~\ref{tab:root_cause_main},\ref{tab:solution_main}, and~\ref{tab:evidence_main}. 
\vspace{-1mm}
\subsubsection{w/o LLM Expert Agents}
We see a drastic drop in all metrics, such as $+8.71$ to $+3.16$ METEOR on root cause prediction, leaving only marginal improvement over ReAct. 
This shows the power of building analytical tools for the LLM, relieving the burden of the controller agent directly analyzing complex data. 
\vspace{-1mm}
\subsubsection{w/o JsonRegen}
The controller and expert agents generate more malformed output, and RCAgent loses a large proportion of its performance, primarily due to erroneous decisions. 
\vspace{-1mm}
\subsubsection{w/o OBSK}
The controller agent cannot use snapshots anymore and has to operate on truncated data. 
The absence of snapshots impacts the overall metrics, including $-1.90$ BLEURT and $-0.69$ G-Correctness on root cause prediction, though not as dramatically as excluding the LLM experts. 
This indicates that the controller can still put analysis on the log with the expert, while a large part of the environmental observation is lost. 

Additionally, we test altering the observation head shown to the controller agent. 
When the observation head is removed and only a snapshot is shown~(w/o Obs Head), the performance slightly downgrades for $-1.10$ BLEURT, implying the major benefit of the snapshot mechanism. 
When we prompt the LLM to summarize the original observation as the alternative head, minimal performance degradation is observed including $-0.06$ G-Correctness. 
The difference except for BLEURT shows that the truncated observation helps more than the summarized content for the controller agent.

\vspace{-1mm}
\input{table/tracestatistics}
\subsection{Stability}
We study the action trajectories of different settings in Table~\ref{tab:tracestatistics}. 
With all enhancements, the RCAgent achieves a $99.38\%$ Pass Rate within 15 steps and a $7.93\%$ Invalid Rate, meaning nearly perfect stability and a significant edge over ReAct. 
With such a minuscule chance of generating problematic actions, RCAgent consistently delivers more accurate and helpful RCA results with shorter trajectories. 

When the LLM expert or OBSK is removed, the controller agent maintains a Pass Rate exceeding $90\%$ while both absences lead to an error-prone exploration, diverting some of its actions toward redundant tool invocations.
The removal of JsonRegen significantly damages the stability due to surging invalid data interchange. 

Moreover, when the default decoding strategy for the controller agent is changed to nucleus sampling~(w/ Sampling), the stability collapses to $70.19\%$ Pass Rate and $44.80\%$ Invalid Rate with tons of erroneous actions and hallucinations. 
Such results highlight the vitality of optimal decoding during initial steps and lead to the design of our mid-way TSC rather than a pure full-process SC. 

We also test using the SQL and SLS query execution tools as the information-gathering tool set. 
A thorough description of all relevant databases is present in the prompt. 
The substitution drastically downgrades the Pass Rate by $33.54\%$, and the gap mainly results from a $70.95\%$ Invalid Rate. 
This clearly shows the necessity of providing semantically minimalist tools for locally hosted LLMs. 

\begin{figure}[!tb]
\centering
\hfill \vspace{-3mm}
  \begin{subfigure}{.22\textwidth}
  \centering
  \includegraphics[width=\linewidth]{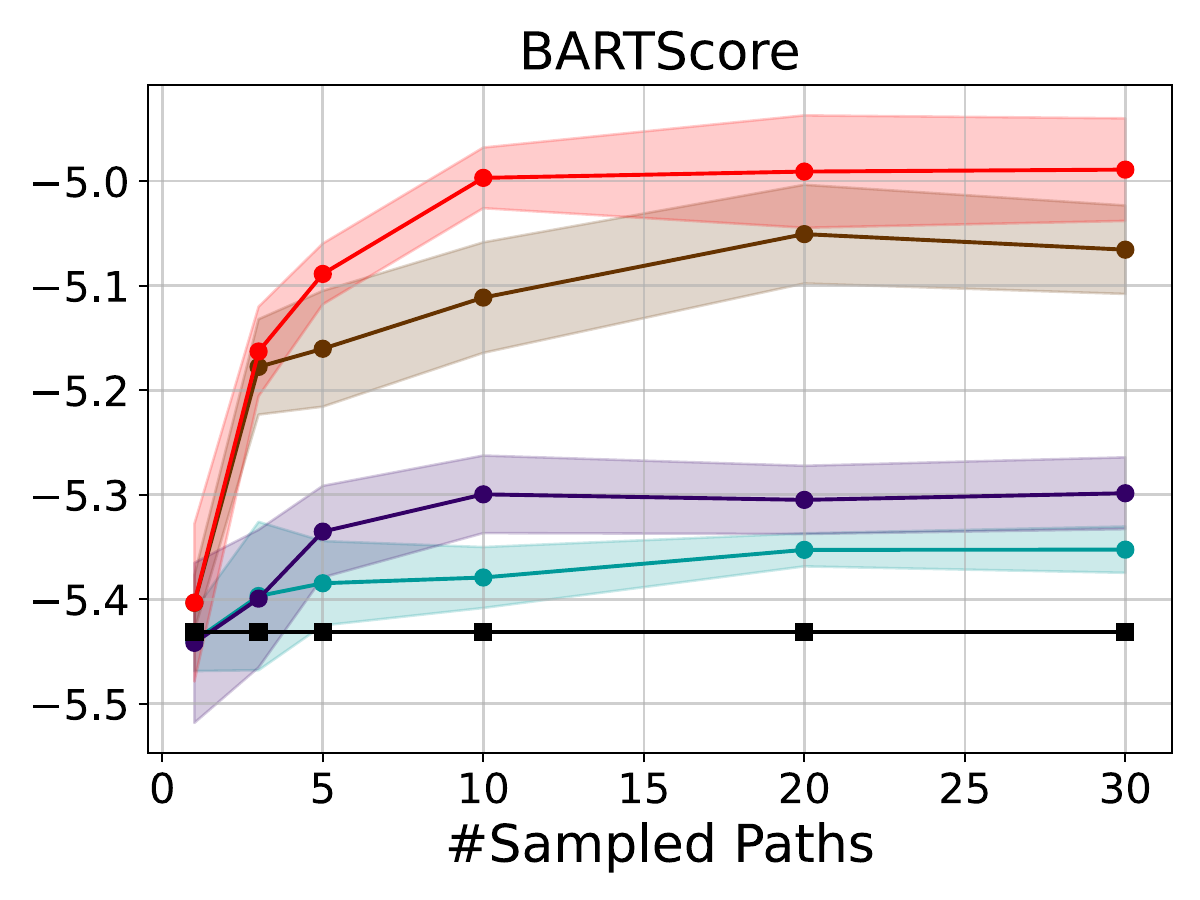}
  \end{subfigure}%
\hfill
  \begin{subfigure}{.22\textwidth}
  \centering
  \includegraphics[width=\linewidth]{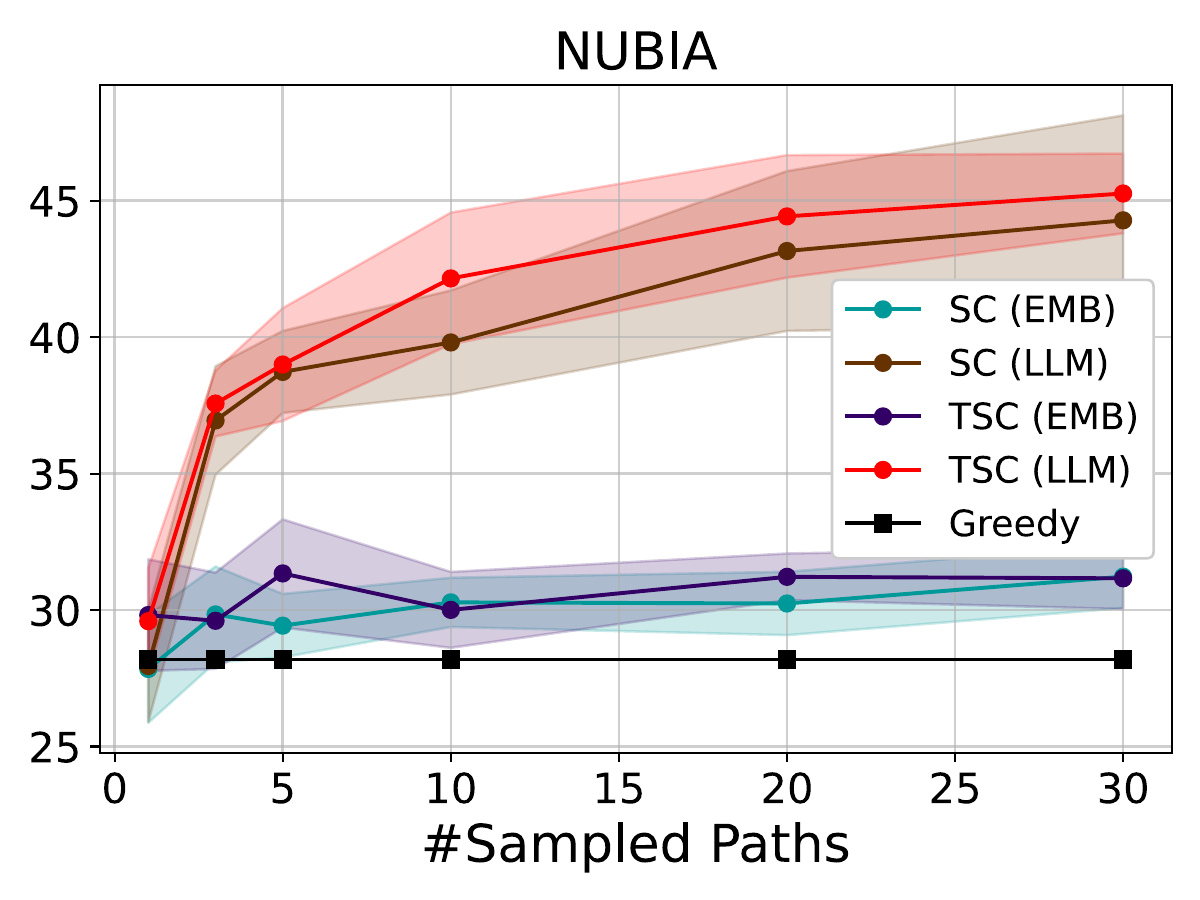}
  \end{subfigure}%
\hfill \vspace{-1mm}
  \caption{
  \label{fig:sc_result}
     Performance of Self-Consistency at different scales and methods. 
     The solid line is the mean score, and the shade represents the standard deviation. 
     The score is calculated on the concatenated solution and root cause. 
  }\vspace{-4mm}
\end{figure}

\subsection{Self-Consistency }
\label{exp:sc}

We study combinations of SC methods and sample counts each for $10$ different runs. The results are detailed in Figure~\ref{fig:sc_result}. 
We evaluate concatenated predicted root causes and solutions for better numerical readability. 

The statistics show that every SC method consistently augments the performance of RCAgent in terms of BARTScore and NUBIA.
This enhancement seems to plateau when the number of samples reaches $20$. 
Among different methods, TSC brings superiority due to its diverse action sampling. 
In all metrics, LLM aggregation outperforms embedding voting, and this gap broadens with an increasing number of samples, illustrating LLM aggregation's ability to offer more comprehensive results as the candidate pool grows. 

\input{table/realworld}

\begin{figure*}[!tb]
    \centering
\includegraphics[width=0.7\linewidth]{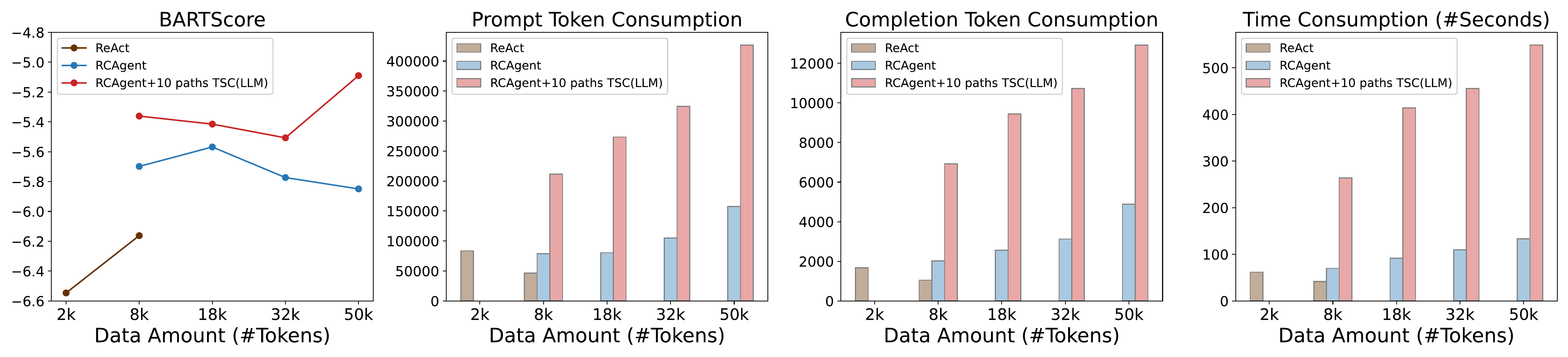}
\vspace{-3mm}
    \caption{
    Performance and resource consumption at different data scales. 
    }
    \label{fig:data_amount}
    \vspace{-3mm}
\end{figure*}

\section{Deployment}
RCAgent has been integrated as a feedback mechanism into the internal operations management platform in our company, aiding in the detection of potential platform-side errors and bugs.
Specifically, RCAgent analyzes all Out-of-Domain~(OoD) jobs that existing automatic SRE tools cannot properly handle. 
Then, jobs labeled as platform responsibility by RCAgent, along with the RCA results, are handed to the SRE team of Apache Flink to aid human diagnosis. 
This procedure improves the efficiency of spotting anomalies that require the DevOps team's attention and the workflow of the after-sale service team. 

\subsection{Performance on OoD Jobs}

We evaluate the performance of RCAgent on OoD jobs in the first two weeks of the system's deployment. 
These OoD job samples are semantically clustered, all of which are beyond the capability of current rules and are labeled by experienced SREs for system evaluation. 
The SRE team is instructed to assign a $0$-$5$ \textbf{H-Helpfulness} score with a set of Likert Scales, spanning from misleading to exceptionally helpful diagnosis. 

The result is shown in \ref{tab:realworld}. Consistent with all LLM and semantic metrics, human evaluators rate our method with an H-Helpfulness of $2.92$, indicating moderate support for RCA. A Tukey's HSD test shows that our method with~($t=5.84, p=0.001$) and without~($t=4.08, p=0.001$) TSC substantially outperforms ReAct. 
Furthermore, equipped with TSC~(LLM), RCAgent demonstrates a precision of $82.06\%$ in determining the responsibility. 

We also examine non-agent RCA solutions for comparison. 
We use all possible types of relevant data of a job, truncated if exceeding the model length constraint, to train XGBoost using document embeddings, fine-tune T5, and feed to LLM with a summarization instruction. 
Despite all the instabilities an autonomous agent solution would incur, which might explain ReAct being worse than traditional methods, our design outperforms non-agent approaches, underscoring the capability of LLM's autonomous decision-making. 

\vspace{-1mm}
\subsection{Computational Scalability}
We show the statistics of LLM agent performance and resource consumption against varying amounts of data included for different jobs during our deployment in Figure~\ref{fig:data_amount}. 
The data amount is calculated by tokenizing all data either presented to the controller agent or queried by the OBSK mechanism. 
While utilizing more data and consuming more resources, RCAgent and its TSC enhancement clearly show better performance compared to ReAct. 
Moreover, our methods' resource consumption scales nearly linearly as the data amount grows~(Pearson Correlation p-value < $0.05$ for any type of consumption), without substantial performance degradation~($p = 1.0$ based on Kruskal-Wallis H-test). 
This breakdown of performance and consumption shows RCAgent's effectiveness in handling a large amount of data.

\section{Related Work}
\subsection{LLM as Autonomous Agents}

As LLMs demonstrate impressive capabilities~\cite{openai2023gpt4, openaichatgptblog, wei2022emergent}, some literatures~\cite{park2023generative, sumers2023cognitive, xi2023rise} leverage these models to construct LLM-based agents.
A popular paradigm is autonomous agents~\cite{wang2023interactive, 2023autogpt, wang2023survey,park2023generative,2023babyagi}, in which LLM agents explore self-directedly without human intervention and step-by-step instructions. 
They perform tasks with an action trajectory, adapting their intentions and outputs according to the environment~\cite{liu2023training}. 
While autonomous agents have been implemented and tested on a variety of toy tasks~\cite{liu2023agentbench, zhou2023llm}, 
our RCAgent is the first work to introduce autonomous LLM agents to realistic cloud RCA tasks.

\subsection{LLM Augmented by Tools}
Recent studies~\cite{qin2023toolllm, vemprala2023chatgpt, qin2023tool, schick2023toolformer} have showcased the proficiency of LLMs to invoke tools and make decisions across a wide range of tasks. 
The tools, in the form of simple functions or external APIs, extend LLM's knowledge and capability evidently~\cite{bubeck2023sparks,ruan2023tptu,li2023api}. 
While stronger LLM can easily grasp tools and accomplish tasks, others can be taught by generated and filtered trajectories~\cite{schick2023toolformer,qin2023toolllm}. 
In this paper, we aim to augment agents with a comprehensive toolset, extending the tool-using paradigm to the real-world cloud RCA. 

\subsection{Cloud RCA with LLMs}
RCA in large cloud services is a prominent subject of study within software engineering communities~\cite{chen2019understanding, ghosh2022fight}. 
A large part of RCA is coupled with NLP due to subtasks like log analysis~\cite{locke2021logassist,le2022log,guo2021logbert}. 
As LLMs advance, they are leveraged for cloud RCA tasks with fine-tuning~\cite{jin2023assess,ahmed2023recommending} or in-context learning~\cite{chen2023empowering,jiang2023xpert}. 
However, these models are not aware of the workflow of cloud RCA, leaving them simply analytical tools. 
We thus investigate tool-augmented LLM as agents for the ever-changing environment of cloud RCA. 

\section{Conclusion}
In this work, we introduce RCAgent, a tool-augmented LLM autonomous agent tailored for cloud root cause analysis. 
RCAgent ensures secure industrial usage of LLM agents in cloud systems by utilizing internally deployed models instead of powerful external ones like ChatGPT. 
Our methodology encompasses a spectrum of enhancements including unique Self-Consistency for action trajectories, a comprehensive prompting framework, expert agents, and stabilization methods. 
Furthermore, RCAgent's efficacy is demonstrated by its practical application in the Real-time Compute Platform for Apache Flink of Alibaba. 
In general, this work pioneers the real-world application of LLM agents in the cloud RCA field.

\bibliographystyle{ACM-Reference-Format}
\balance
\bibliography{ref}


\end{document}

%% file: table/rootcause.tex

\begin{table}[!tb]

\centering
\footnotesize

\caption{
    \label{tab:root_cause_main}
    Results of root cause prediction on Flink jobs. 
}
\vspace{-3mm}
\resizebox{0.50\textwidth}{!}{
\begin{tabular}{l | c c  c c  | c  }
\toprule
\multicolumn{1}{l|}{\multirow{3}*{Model}}
& \multicolumn{4}{c|}{\bf Semantic Metrics} 
& \multicolumn{1}{c}{\bf LLM Metric} \\
\cmidrule(lr){2-5} \cmidrule(lr){6-6} 
 & \multicolumn{1}{c}{METEOR}  
 & \multicolumn{1}{c}{EmbScore} 
 & \multicolumn{1}{c}{BLEURT} 
 & \multicolumn{1}{c|}{BARTScore} 
 & \multicolumn{1}{c}{G-Correctness} 
 \\ 
\midrule


ReAct
& 6.44 
& 89.64
& 25.17 
& -6.20
& 3.06

\\
RCAgent
& 15.15 
& 91.47
& 31.57 
& -5.74
& 5.22
\\
\quad w/o LLM experts
& 9.60 
& 90.33
& 27.77
& -6.02 
& 3.97
\\
\quad w/o JsonRegen
& 13.89 
& 90.74
& 27.72 
& -5.84
& 4.19
\\
\quad w/o OBSK
& 12.37 
& 90.97
& 29.67 
& -5.77
& 4.53
\\
\quad w/o Obs Head
& 12.27 
& 91.30
& 30.47 
& -5.87
& 4.98
\\
\quad w/ Summary Head
& 14.64
& 91.34
& 32.11 
& -5.82
& 5.16
\\
\midrule


\quad w/ SC~(LLM)
& 15.94\textsubscript{\textpm0.44}
& 91.59\textsubscript{\textpm0.06}
&  33.74\textsubscript{\textpm0.44}
& -5.48\textsubscript{\textpm0.04}
& 5.38\textsubscript{\textpm0.01}
\\
\quad w/ TSC~(LLM)
& \bf 16.49\textsubscript{\textpm0.09}
& \bf 91.67\textsubscript{\textpm0.03}
& \bf 34.43\textsubscript{\textpm0.59}
& \bf -5.40\textsubscript{\textpm0.01}
& \bf 5.47\textsubscript{\textpm0.06}
\\

\bottomrule

\end{tabular}

}
\vspace{-3mm}
\end{table}

%% file: table/solution.tex

\begin{table}[tb!]

\centering
\footnotesize

\caption{
    \label{tab:solution_main}
    Results of solution generation on Flink jobs. 
}\vspace{-3mm}
\begin{tabular}{l | c c c | c }
\toprule
\multicolumn{1}{l|}{\multirow{3}*{Model}} 
& \multicolumn{3}{c|}{\bf Semantic Metrics} & \multicolumn{1}{c}{\bf LLM Metric} \\
\cmidrule(lr){2-4} \cmidrule(lr){5-5} 
 & METEOR  & BLEURT & BARTScore & G-Helpfulness \\ 
\midrule
ReAct
 & 6.42 
 & 26.97 
 & -4.90 
 & 3.41
 \\
RCAgent
& 12.94 
& 34.68 
& -4.17 
& 5.48
\\
\quad w/o LLM experts
& 8.46 
& 30.46 
& -4.63
& 4.13
\\ 
\quad w/o JsonRegen
& 11.41 
& 31.25 
& -4.40
& 4.58
\\ 
\quad w/o OBSK
 & 10.34 
 & 32.13 
 & -4.37 
 & 4.68
 \\ 

\midrule


\quad w/ SC~(LLM)
& 15.27\textsubscript{\textpm0.19}
& 37.94\textsubscript{\textpm0.03}
& -4.00\textsubscript{\textpm0.00}
& 5.55\textsubscript{\textpm0.03}
\\
\quad w/ TSC~(LLM)
& \bf 16.45\textsubscript{\textpm0.06} 
& \bf 39.18\textsubscript{\textpm0.13}
& \bf -3.94\textsubscript{\textpm0.08}
& \bf 5.69\textsubscript{\textpm0.02} 
\\

\bottomrule

\end{tabular}

\vspace{-2mm}
\end{table}

%% file: table/evidence.tex

\begin{table}[!htbp]

\footnotesize
\vspace{-3mm}
\caption{
    \label{tab:evidence_main}
    Semantic scores of evidence of methods. 
}
\vspace{-3mm}
\begin{tabular}{l | c c c  }
\toprule
\multicolumn{1}{l|}{\multirow{3}*{Model}} 
& \multicolumn{3}{c}{\bf Semantic Metrics} \\
\cmidrule(lr){2-4} 
 & METEOR  & EmbScore  & BARTScore \\ 
\midrule
ReAct
 & 11.82 & 90.03 & -5.74 \\
RCAgent
& 28.10 & 92.14 & -4.62 \\
\quad w/o LLM experts
& 13.10 & 90.63 & -5.63 \\
\quad w/o OBSK
 & 17.79 & 91.12 & -5.13 \\ 
\quad w/ Summary Head 
& 18.09& 91.67&-5.14 \\

\midrule


\quad w/ SC~(LLM)
& 30.15\textsubscript{\textpm0.83} & 92.60\textsubscript{\textpm0.06} & -4.41\textsubscript{\textpm0.05}\\
\quad w/ TSC~(LLM)
& \textbf{30.84\textsubscript{\textpm0.43}} & \textbf{92.78\textsubscript{\textpm0.02}} & \textbf{-4.29\textsubscript{\textpm0.02}} \\

\bottomrule

\end{tabular}

\vspace{-3mm}
\end{table}

%% file: table/tracestatistics.tex
\begin{table}[tbp!]

\centering
\footnotesize

\caption{
    \label{tab:tracestatistics}
    Trajectory statistics of different settings. 
    }
\vspace{-3mm}
\begin{tabular}{l | c c c }
\toprule 
 Model &  Pass Rate  & Trajectory Length  
 &  Invalid Rate \\ 
\midrule
ReAct
& 86.33 
& 7.48
& 22.82  \\
RCAgent
& \bf 99.38 
& \bf 6.78 
& \bf 7.93 \\
\quad w/o LLM experts
& 92.55 
& 6.93
& 16.24 \\ 
\quad w/o JsonRegen
& 85.71 
& 7.91
& 18.75 \\ 
\quad w/o OBSK
& 96.89 
& 7.21
& 18.34 \\ 

\midrule
\quad w/ Sampling
& 70.19
& 10.66
& 44.80  \\
\quad w/ SQL tools
& 65.84
& 10.55
& 70.94 \\



\bottomrule

\end{tabular}
\vspace{-3mm}
\end{table}

%% file: table/realworld.tex

\begin{table*}[!tb]

\centering
\footnotesize

\caption{
    \label{tab:realworld}
    Evaluations on the online OoD anomalies. \textbf{Bold} denotes the best results. 
}
\vspace{-3mm}
\begin{tabular}{l | c c c c c   | c | c }
\toprule
\multicolumn{1}{l|}{\multirow{3}*{Model}}
& \multicolumn{5}{c|}{\bf Root Cause} 
& \multicolumn{1}{c|}{\bf Responsibility}
& \multicolumn{1}{c}{\bf Human}\\
\cmidrule(lr){2-6} \cmidrule(lr){7-7} \cmidrule(lr){8-8} 
 & \multicolumn{1}{c}{METEOR}  
 & \multicolumn{1}{c}{NUBIA}
 & \multicolumn{1}{c}{BLEURT}  
  & \multicolumn{1}{c}{BARTScore}
 & \multicolumn{1}{c|}{G-Correctness} 
 & \multicolumn{1}{c|}{Precision} 
 & \multicolumn{1}{c}{H-Helpfulness} 
\\
\midrule

XGBoost
&  - 
& -
&  -
&  -
&  -
& 77.65
&  -
\\


Finetune T5
&  4.18
& 8.20
&  19.32
&  -6.61
&  2.62
&  77.85
&  -
\\

LLM Summary
&  7.88
& 14.57
&  25.40
&  -6.00
&  3.58
& 77.21
&  -
\\

\midrule
ReAct
&  5.21 
& 10.38
&  20.33
&  -6.25
& 2.24
& 73.53
&  1.36\textsubscript{\textpm0.03} 

\\

RCAgent
&  13.77 
& 19.48
&  31.52
&  -5.59 
& 3.82
& 80.74
&  2.47\textsubscript{\textpm0.17} 

\\


\quad w/ TSC~(LLM)
& \bf 15.72\textsubscript{\textpm0.61} 
& \bf 26.79\textsubscript{\textpm2.54}
& \bf 35.72\textsubscript{\textpm0.58}
& \bf -5.29\textsubscript{\textpm0.03}
& \bf 4.36\textsubscript{\textpm0.01}
& \bf 82.06\textsubscript{\textpm0.42}
& \bf 2.92\textsubscript{\textpm0.21} 

\\

\bottomrule

\end{tabular}

\vspace{-3mm}
\end{table*}